\documentclass[11pt,epsf,letterpaper,fleqn]{article}
\usepackage{float}
\usepackage{setspace}
\usepackage{caption}
\usepackage{color}
\usepackage{amsmath}
\usepackage{amsfonts}
\usepackage{subcaption}
\usepackage{comment}
\usepackage{booktabs}
\usepackage{verbatim}
\usepackage{amssymb}
\usepackage{cite}
\usepackage{graphicx}
\usepackage{amsmath}
\usepackage{xcolor}
\usepackage{physics}

\providecommand{\U}[1]{\protect\rule{.1in}{.1in}}
\begin{document}

\title{Rotationally invariant Isospinning Baby-Skyrmions dressed by Fermions}
\author{Marco Barsanti$%
^{a} $, Gianni Tallarita$^{b}$ \\
\vspace{.1 cm}\\
{\normalsize \textit{$^{a}$Department of Physics “E. Fermi”, University of Pisa and INFN, Sezione di Pisa}}\\
{\normalsize \textit{Largo Pontecorvo, 3, Ed. C, 56127 Pisa, Italy. \vspace{.4 cm}}}\\
{\normalsize \textit{$^{b}$Departamento de Ciencias, Facultad de Artes
Liberales, Universidad Adolfo Ib\'a\~nez,}}\\
{\normalsize \textit{Santiago 7941169, Chile. \vspace{.4 cm}}}\\
{\footnotesize marco.barsanti@phd.unipi.it, gianni.tallarita@uai.cl.} }
\date{}
\maketitle
\begin{abstract}
We couple Fermions to the isospinning (2+1) baby-skyrme model. We show that consistent rotationally invariant localised solutions can be found but the Fermionic solutions to the equations of motion are not in general eigenstates of the Hamiltonian. These do however become full eigenstates in a particular limit. We also find the corresponding Fermionic eigenstates of the Hamiltonian, for which our Fermion ansatz is also well suited. For localised Fermionic states, a novel constraint on the internal isospin frequency exists. Throughout our studies, we include the backreaction of the Fermions on the baby-skyrmion.
\end{abstract}

\tableofcontents

\vspace{.1 cm}

\hfill

\section{Introduction}

\indent The Skyrme model \cite{Skyrme:1961vq} has long been the subject of intense study (for reviews see \cite{Manton:2004tk} \cite{Shifman:2012zz}  and references therein).  The reason is its intimate relation to low energy Baryons in the limit of large number of colour charges \cite{Witten:1983tw}. The model can be thought to arise as an effective field theory from the integration of high energy Fermionic ``quark" modes. In this sense the theory does not intrinsically possess any fermionic degrees of freedom. While phenomenologically appealing, the full Skyrme theory is not particularly easy to study. It is a highly non-linear theory in four dimensions and  most theoretical progress has to be made numerically. While being a simpler model overall, the baby-Skyrmion model \cite{Piette:1994ug} \cite{Bogolyubskaya:1989fz} \cite{Bogolubskaya:1989ha} captures the essence and main features of its higher dimensional parent remarkably well. This lower dimensional model is particularly suited for the study of Skyrmion interactions in condensed matter systems \cite{cond1} \cite{cond2} \cite{cond3} and more recently for quantum computation purposes \cite{comp1}  \cite{comp2} \cite{comp3}. This model also yields solitonic solutions with Baryon charge but, unlike the parent theory, it is not a-priori built from integrating Fermionic degrees of freedom and therefore can be naturally coupled to them.\newline

The static baby-skyrmion solutions admit an isospinning generalization  \cite{Halavanau:2013vsa}  \cite{Battye:2013tka}  \cite{Shnir:2014yhj} \cite{Harland:2013uk}. This endows the baby-skyrmion with time dependence, in the form of rotations in internal space parametrised by a frequency $\omega$. This frequency is severly restricted: it must obey positivity of the effective kinetic terms in the Lagrangian and be such that perturbations of the effective potential are not tachyonic. The presence of isospin provides the baby-skyrmion with a new instability channel. In particular, for Baryon numbers $B\geq 2$, the internal rotation can cause a decay into $B=1$ constituent baby-skyrmion solutions. The rotationally symmetric $B=1$ solutions are stable under such isospinning rotations.  \newline

 The topic of Fermion couplings to baby-Skyrmions is a very interesting one, neatly parametrized by the Fermion-baby-Skyrmion coupling constant $g$. Most of the theoretical work regarding this coupled system is performed in the decoupling limit of $g\rightarrow 0$ \cite{Kodama:2008xm} \cite{Delsate:2011aa} , where one can safely assume the Fermions do not modify the baby-Skyrmion profile. Fermion scatterings on baby-Skyrmion solutions were previously investigated in \cite{Loginov:2021rka}. Supersymmetric extensions of this system were also previously considered \cite{Adam:2011hj}. Recently however, back reaction of the Fermions was included by raising the value of $g$ and solving the resulting coupled system numerically \cite{Perapechka:2018yux} \cite{Perapechka:2019upv}. This was done for $B=1$ baby-skyrmion solutions with no internal isospin, the results show a continuous space of Fermion eigenstates coupled to the baby-skyrmion.  \\

 In this paper we wish to extend this analysis to include internal isospin. Restricting to $B=1$ rotationally symmetric stable baby-skyrmion solutions, we wish to investigate how one can couple these to Fermions. Unlike the previous case where the Fermions are eigenstates of the resulting Hamiltonian, we will show that  the inclusion of isospin means such Fermions are no longer generically eigenstates, becoming so only in a very particular limit.  We will therefore interpret the resulting energy of the system as an expectation value of the mean energy of the system, and not as an energy eigenvalue of the Hamiltonian. Throughout our studies, we include the backreaction on the baby-skyrmion.   \newline

The paper is structured as follows: in section \ref{due} we introduce the action of the baby-skyrme system coupled with Fermions, section \ref{sect2} is dedicated to the solutions of this system and finally in section \ref{sectconc} we provide the conclusions of our research and some avenues of future investigation.

\section{The System}
\label{due}
 The action we use describing a baby Skyrme
system coupled to a Fermion is 
\begin{eqnarray}\label{action}
S_{{}} &=&\int d^{3}x\,\Big[\frac{a_{1}}{2}\big(\nabla _{\mu }\vec{%
\Phi}\big)\cdot \big(\nabla ^{\mu }\,\vec{\Phi}\big)-\kappa _{0}\big(1-%
\vec{\Phi}\cdot \vec{n}\big)\Big.  \notag  \label{skyrmaction} \\
&&\quad \qquad \Big.-\frac{a_{2}}{4}\big(\nabla _{\mu }\vec{\Phi}\times
\nabla _{\nu }\vec{\Phi}\big)\cdot \big(\nabla ^{\mu }\,\vec{\Phi}\times
\nabla ^{\nu }\,\vec{\Phi}\big)%
+\mathcal{L}_{F}\Big]\;,
\end{eqnarray}%
\begin{equation}
\mathcal{L}_{F}=\overline{\Psi }\left( i\gamma_{\mu }\nabla ^{\mu }+g\,\vec{
\tau }\cdot \vec{\Phi}-m\right) \Psi \;,  \label{def}
\end{equation}%
where $g$ is the
coupling constant related to the baby-Skyrmion-Fermion interaction and the $a_1$, $a_2$ are constant coefficients. $\vec{\Phi}$ is a vector field of dimension $n=3$. The isospin matrices are defined as $\tau^i=\mathbb{I}\otimes \sigma^i$, where $\sigma^i$ are the Pauli matrices. The spin matrices are defined as $\gamma_{\mu}=\tilde{\gamma}_{\mu}\otimes\mathbb{I}$. We also include a potential term for the baby-Skyrmion with its corresponding coupling constant $\kappa_0$. The general literature also considers different kinds of potentials given by higher powers of the one chosen above. In this paper, we restrict our attention to this potential only.

 The mass dimensions of the parameters appearing in the Lagrangian are
\begin{equation}
[g]=[m]=[a_1] = +1, \quad [a_2] = -1,\quad [\kappa_0] = +3.
\end{equation}
In this work, we use the flat metric $\eta^{\mu\nu}=(1,-1,-1)$ and we choose the spin matrices $\tilde{\gamma}_1=-i\sigma_1$, $\tilde{\gamma}_2=-i \sigma_2$ and $\tilde{\gamma}_0=\sigma_3$ such that $\gamma^{\mu}$ respect the Clifford algebra
\begin{equation}
    \{\gamma^{\mu},\gamma^{\nu}\}=2\mathbb{I}\,\eta^{\mu\nu}.
\end{equation}
For the baby-Skyrmion, we will use the following parametrization 
\begin{eqnarray}
&&\vec{\Phi}\cdot \vec{\Phi} =1\quad \Longleftrightarrow \quad \vec{\Phi}%
=\left( \sin F\cos G,\sin F\sin G,\cos F\right) \;,  \label{def2} \\
&&F =F\left( x^{\mu }\right) \;,\quad G=G\left( x^{\mu }\right) \ ,
\end{eqnarray}where $F$ and $G$ are real-scalar fields.
On the other side, for the fermion field we choose
\begin{equation}
\Psi =1/N_{i}\left( 
\begin{array}{c}
v_{1} \exp \left[ i \Omega _1 \right]
\\ 
v_{2} \exp \left[ i\Omega _2 \right] \\ 
u_{1} \exp \left[ i\Omega _3\right] \\ 
u_{2} \exp \left[ i\Omega _4 \right]%
\end{array}%
\right)\,,  \label{fermionansatz11}
\end{equation}where $v_{1}=v_{1}\left( x^{\mu }\right)$, $v_{2}=v_{2}\left( x^{\mu }\right)$, $u_{1}=u_{1}\left( x^{\mu }\right)$, $u_{2}=u_{2}\left( x^{\mu }\right)$, $\Omega_{1}=\Omega_{1}\left( x^{\mu }\right)$, $\Omega_{2}=\Omega_{2}\left( x^{\mu }\right)$, $\Omega_{3}=\Omega_{3}\left( x^{\mu }\right)$ and  $\Omega_{4}=\Omega_{4}\left( x^{\mu }\right)$ are real-scalar field components. The coefficient $N_i$ is a normalization coefficient which ensures that $\int \Psi^\dagger\Psi\; d^2x= 1$. In detail, this evaluates to
\begin{equation}\label{intcont}
N_i^2 =  \int\left(u^*_1u_1+u_2^{*}u_2+v_1^{*}v_1+v_2^{*}v_2\right) d^2x.
\end{equation}The field $\Psi$ is a spin-isospin spinor, meaning that it describes the wave function of a particle with spin one-half and two possible state of iso-spin: up and down.

Without loss of generality, we also take the standard choice for the vacuum 
\begin{equation}
\vec{n}=\left( 0,0,1\right) \;.  \label{def3}
\end{equation}

Given our choice for the metric, the positivity of the energy of the baby skyrmion sector is guaranteed requiring
\begin{equation}
    a_1>0,\quad a_2\geq 0,\quad \kappa_0\geq 0.
\end{equation}
Finally, the topological charge of the baby Skyrmion is%
\begin{equation}
B=\frac{1}{4\pi }\int_{\Sigma }\rho _{B}\;, \quad \rho _{B}=\frac{1}{4\pi }%
\left( \sin F\right) dF\wedge dG\;.  \label{rational4.1.1}
\end{equation}%

The equations of motion for the baby skyrmion fields are%
\begin{eqnarray}
0 &=&-\square F+\frac{\sin \left( 2F\right) }{2}\nabla _{\mu }G\nabla ^{\mu
}G  \notag \\
&&+c_{1}\frac{\sin (2F)}{2}\left[ (\nabla _{\mu }F\nabla ^{\mu }F)(\nabla
_{\nu }G\nabla ^{\nu }G)\right. -\left. (\nabla _{\mu }F\nabla ^{\mu }G)^{2}%
\right]  \notag \\
&&-c_{1}\nabla _{\mu }\left[ \sin ^{2}(F)\left[ (\nabla _{\nu }G\nabla ^{\nu
}G)\nabla ^{\mu }F-(\nabla _{\nu }F\nabla ^{\nu }G)\nabla ^{\mu }G\right] %
\right]  \notag \\
&&+\frac{\kappa_0}{a_{1}}\frac{\partial }{\partial F}\big(1-\cos{F}\big)\ -\frac{g}{a_1}\frac{\delta }{\delta F}%
\overline{\Psi }\left( \overrightarrow{\tau }\cdot \vec{\Phi}\right) \Psi
\label{equF1}
\end{eqnarray}%
\begin{eqnarray}
0 &=&-\sin ^{2}(F)\square G-\sin \left( 2F\right) \nabla _{\mu }F\nabla
^{\mu }G-\frac{g}{a_1}\frac{\delta }{\delta G}\overline{\Psi }\left( \overrightarrow{%
\tau }\cdot \vec{\Phi}\right) \Psi  \notag \\
&&+c_{1}\nabla _{\mu }\left[ \sin ^{2}(F)\left[ (\nabla _{\nu }F\nabla ^{\nu
}G)\nabla ^{\mu }F(\nabla _{\nu }F\nabla ^{\nu }F)\nabla ^{\mu }G\right] %
\right] \;.  \label{equG2}
\end{eqnarray}%
In the above equations $c_{1}=\frac{a_{2}}{a_{1}}$. Similarly, the Dirac equations for the Fermions read%
\begin{equation}
\left( i\gamma ^{\mu }\nabla _{\mu }+g\,\vec{\tau }\cdot \vec{\Phi}%
-m\right) \Psi =0.  \label{Dirac}
\end{equation}

\subsection{Symmetries}

The system admits at least two internal exact symmetries. The first symmetry of the action (\ref{action}) is an $SO(2)\sim U(1)$ rotation that acts only on the field components $G$, $\Omega_2$ and $\Omega_4$ as
\begin{align}
   & G\rightarrow G'=G+\alpha\\
   & \Omega_2\rightarrow \Omega_2'=\Omega_2+\alpha\\
   &\Omega_4\rightarrow \Omega_4'=\Omega_4+\alpha
\end{align}where $\alpha$ is a constant phase. Equivalently, this rotation acts on the fundamental field $\Phi$ and $\Psi$ as
\begin{equation}
    \Phi\rightarrow\Phi'=O\Phi\qquad \Psi\rightarrow\Psi'=e^{iQ\alpha}\Psi
\end{equation}where
\begin{equation}\label{chargeO}
    O=\begin{pmatrix}
\cos \alpha & -\sin \alpha & 0\\
\sin \alpha & \cos \alpha & 0\\
0&0& 1
\end{pmatrix}\qquad Q=\frac{1}{2}(\mathbb{I}-\tau_3)
\end{equation}and it corresponds to an iso-spin rotation.
The related conserved current $J_{U(1)}^{\mu}=J^{\mu}_s+J^{\mu}_f$ contains the scalar-field term $J_s^{\mu}$
\begin{equation}
J_{s }^{\mu}=\,a_{1}\sin ^{2}F\left( 1-\frac{a_{2}}{a_{1}}\left( F^{\prime
}\right) ^{2}\right) \nabla ^{\mu }G\   \label{scc1}
\end{equation}
and the Fermion current component $J^{\mu}_f$ 
\begin{equation}\label{sccf}
J_{f }^{\mu}= \bar{\Psi}\gamma^\mu Q\Psi.
\end{equation}

The previous definition of the isospin operator $Q$ can help us to better clarify the physical meaning of the spin-isospin spinor $\Psi$ in $2+1$ dimensions. The last ingredient that we need to this end is the definition of the spin operator that, following \cite{Perapechka:2018yux}, reads
\begin{equation}
    S=\frac{\gamma_3}{2}\,.
\end{equation}Now, given the basis
\begin{equation}
\hat{e}_1 =\left( 
\begin{array}{c}
1
\\ 
0 \\ 
0 \\ 
0
\end{array}%
\right)\,, \quad 
\hat{e}_2 =\left( 
\begin{array}{c}
0
\\ 
1 \\ 
0 \\ 
0
\end{array}%
\right)\,, \quad
\hat{e}_3 =\left( 
\begin{array}{c}
0
\\ 
0 \\ 
1 \\ 
0
\end{array}%
\right)\,, \quad
\hat{e}_4 =\left( 
\begin{array}{c}
0
\\ 
0 \\ 
0 \\ 
1
\end{array}%
\right)\,, 
\end{equation}it is a simple computation to show that the isospin and spin operator $Q$ and $S$ act on these vectors as
\begin{equation}
    \begin{aligned}
        & Q\hat{e}_1=0\qquad S\hat{e}_1=1/2\\
        & Q\hat{e}_2=1\qquad S\hat{e}_2=1/2\\
      & Q\hat{e}_3=0\qquad S\hat{e}_3=-1/2\\
           & Q\hat{e}_4=1\qquad S\hat{e}_4=-1/2\,.\\
    \end{aligned}
\end{equation}The spinor $\Psi$ describes therefore a particle state with two possible value of spin and isospin. In this case, the complex functions $v_{1} \exp \left[ i \Omega _1 \right]$, $v_{2} \exp \left[ i \Omega _2 \right]$, $u_{1} \exp \left[ i \Omega _3 \right]$, $u_{2} \exp \left[ i \Omega _4 \right]$ represent the spatial wave-functions associated with each pair of spin-isospin numbers.

Besides the isospin rotation, the system is also invariant under the phase transformation
\begin{eqnarray}
&&\Omega_1\rightarrow \Omega_1'=\Omega_1+\beta\\
&&\Omega_2\rightarrow \Omega_2'=\Omega_2+\beta\\
&&\Omega_3\rightarrow \Omega_3'=\Omega_3+\beta\\
&&\Omega_4\rightarrow \Omega_4'=\Omega_4+\beta
\end{eqnarray}that correspond to $\Psi\rightarrow\Psi'=e^{i\beta}\Psi$. The associated
Fermionic current $J^{\mu }_{F}$ is%
\begin{equation}
J^{\mu }_{F}=\bar{\Psi }\gamma ^{\mu }\Psi \ .  \label{fc1}
\end{equation}The conservation of this current corresponds to the conservation of quantum probability since the time-component of the four-vector (\ref{fc1}) represents the probability density ($\rho_F$ from now)
\begin{equation}\label{fc11}
    \rho_F=J^0_F=\Psi^{\dagger}\Psi.
\end{equation}The divergence of the current (\ref{fc1}) gives therefore the continuity equation
\begin{equation}\label{fc111}
    \frac{\partial\rho_F}{\partial t}+\mathbf{\nabla}\cdot\mathbf{J}_F=0. 
\end{equation}

\section{Combined Skyrmion-Fermion ansatz}\label{sect2}

\subsection{\protect\normalsize The baby-Skyrmion ansatz}
In order to solve the equations of motion for the baby Skyrmion-Fermion system, from this section we will use spatial polar coordinates $\{x,\varphi\}$, where $x$ is the radial coordinate and $\varphi$ is the angular coordinate.

For what concerns the baby-Skyrmion fields, we use the following ansatz
\begin{equation}\label{babyansatz}
F=F\left( x\right) \;,\ \ G=p \varphi+\omega t \;,\ \quad p\in 
\mathbb{Z}
\end{equation}%

Note that this ansatz implies a non trivial periodic dependence on the time coordinate. One of the fundamental aspects of this kind of ansatz is that even though the field profiles depend explicitly on time, the energy density does not, as we will show in section (\ref{sec:energy}). \newline

As shown in \cite{Halavanau:2013vsa}  \cite{Harland:2013uk}  \cite{Shnir:2014yhj}, insertion of this ansatz in the baby-Skyrme lagrangian (ignoring the Fermions for the time being) leads to an effective Lagrangian in which the isospinning contribution causes two main channels of instability. The first is related to the effective kinetic term for the baby-Skyrmion fields
\begin{equation}
K_{eff} = \int xdx d\varphi\big[\frac{1}{2}\left(1-\omega^2\left(1-\phi_3^2\right)\right)\partial_i\vec{\Phi}\cdot\partial_i\vec{\Phi}+\omega^2\left(\vec{\Phi}\cross\partial_i\vec{\Phi}\right)_3\left(\vec{\Phi}\cross\partial_i\vec{\Phi}\right)_3\big].
\end{equation}
This term becomes singular when $\omega = 1$. The second is related to the effective potential for the scalar field,
\begin{equation}
V_{eff} = \kappa_0(1-\phi_3)-\omega^2\left(1-\phi_3^2\right).
\end{equation}
In order to have a real value for the  ``pion mass", and thus an asymptotically decreasing behaviour for the baby skyrmion profile-function, we require the effective mass of baby-Skyrme excitations to be positive. \newline

We define here an effective angular momentum for the baby-Skyrmion as
\begin{equation}
J =2\pi \omega \int xdx\big[\left(1-\phi_3^2\right)\partial_i\vec{\Phi}\cdot\partial_i\vec{\Phi}-\left(\vec{\Phi}\cross\partial_i\vec{\Phi}\right)_3\left(\vec{\Phi}\cross\partial_i\vec{\Phi}\right)_3\big] \label{effJ}.
\end{equation}

The topological
charge density (\ref{rational4.1.1}) with the ansatz (\ref{babyansatz}) simplifies to%
\begin{equation}
\rho _{B}=\frac{p}{4\pi }\sin F\left( \partial _{x}F\right) dx\wedge d\varphi
\;.  \label{ansch}
\end{equation}%
Given the interval for the $x-$coordinate $x\in[0,\infty]$ and the boundary conditions for $F$
\begin{equation}
F\left(\infty\right)=0\qquad F\left(0\right) =\pi\,,  
\end{equation}the baby Baryon charge $B$ is then $B=p$. Since in this work we want to consider the axially symmetric ansatz (\ref{babyansatz}) for the baby-Skyrmion, we reduce our analysis to the case 
\begin{equation*}
    p=\pm 1\,.
\end{equation*}

Using the baby-Skyrmion ansatz, the Lagrangian for the Fermions can be reduced to
\begin{equation}\label{feq}
    \mathcal{L}_F=\Psi^{\dagger}D \Psi
\end{equation}where the operator $D$ has components

\setlength{\mathindent}{1cm} {\normalsize 
\begin{equation}
D=\left[ 
\begin{array}{cccc}
-m+g\cos F +i\partial _{t} & g \sin F e^{ -i\left( p\varphi
+\omega t\right)} & e^{-i\varphi}\left(-\frac{x\partial _{x}-i\partial_ \varphi }{x} \right)& 0 \\ 
g \sin F e^{ i\left( p\varphi +\omega t\right)} & -m-g\cos F
+i\partial _{t} & 0 & e^{-i\varphi}\left(-\frac{x\partial _{x}-i\partial_ \varphi }{x} \right) \\ 
 e^{i\varphi}\left(\frac{x\partial _{x}+i\partial_ \varphi }{x} \right) & 0 & m-g\cos F +i\partial
_{t} & -g\sin F e^{ -i\left( p\varphi +\omega t\right)} \\ 
0 &  e^{i\varphi}\left(\frac{x\partial _{x}+i\partial_ \varphi }{x} \right) & -g\sin F e^{i\left(
p\varphi +\omega t\right)} & m+g\cos F +i\partial _{t}%
\end{array}%
\right].\label{ham}
\end{equation}

\subsection{\protect\normalsize The Fermion ansatz}
In the previous subsection, we chose the common rotational-invariant ansatz for the baby-Skyrmion, requiring the profile function $F$ to be dependent only on the radial variable $x$. Moreover, although the explicit time-dependence of the field $G$, the baby baryon density $\rho_{B}$ with ansatz (\ref{babyansatz}) does not depend on time and thus, in this sense, we can call that configuration a stationary configuration. Considering this choice for baby-Skyrmion sector, in order to solve the full equations of motion that include the coupling between the baby-Skyrmion and Fermion, we propose an ansatz for the fermion field that is consistent with axial-invariance and stationarity. With this aim, we choose
\begin{equation}
\Psi =1/N_{i}\left( 
\begin{array}{c}
v_{1}\left( x\right) \exp \left[ i \Omega _1(\varphi,t) \right]
\\ 
v_{2}\left( x\right) \exp \left[ i\Omega _2(\varphi,t) \right] \\ 
u_{1}\left( x\right) \exp \left[ i\Omega _3(\varphi,t) \right] \\ 
u_{2}\left( x\right) \exp \left[ i\Omega _4(\varphi,t) \right]%
\end{array}%
\right)  \label{fermionansatz1}
\end{equation}%
with
\begin{equation}\label{ansatzzf}
\begin{aligned}
&\Omega_1=l\; \varphi+k_1 \omega \;t\\
& \Omega_2=(p+l)\;\varphi+(1+k_1) \omega\;t\\
& \Omega_3=(l+1)\;\varphi+k_1\omega t\\
& \Omega_4=(l+p+1)\;\varphi+(1+k_1)\omega t \,,
\end{aligned}
\end{equation}
where the real functions $v_1$, $v_2$, $u_1$ and $u_2$ depend only on the radial coordinate $x$ and in which $l\in \mathbb{Z}$ and $k_1\in \mathbb{R}$. This particular ansatz leads the Fermion density $\rho_F=\Psi^{\dagger}\Psi$ to be rotational-invariant and time-independent, as requested before. In particular, as we will show in the next subsection, it decouples the angular dependence from the equations, leaving radial equations for the field profile functions only. Note, obviously, that the time-dependence of (\ref{ansatzzf}) is different with respect to the case of \cite{Perapechka:2018yux}, in which zero isospining-angular-velocity $\omega$ was considered. This fact leads to a fundamental difference between the two cases: in \cite{Perapechka:2018yux}, a fermionic solution of the equations of motion with rotational-invariance symmetry (and stationary fermionic density) is also simultaneously an Eigenstate of the Hamiltonian; contrarily, as we will show in details in section \ref{energyeigen}, in this case the field (\ref{fermionansatz1}) cannot possess this property. A solution to the equations of motion below is therefore not generically an Eigenstate of the Hamiltonian. It becomes so only in a particular limit. A natural question would therefore be to relax the assumption of time-independence of the the baby-Skyrmion and Fermion profile functions, along with relaxing the need for cylindrical symmetry, and solve the full set of PDE equations. In that case, different considerations about the properties of such a Fermion state would be needed and we leave this for a future work.

\subsection{Equations of motion}\label{eoms}
Using the ansatz (\ref{babyansatz}) and (\ref{fermionansatz1})-(\ref{ansatzzf}), the equations of motion (\ref{equF1})-(\ref{equG2}) and (\ref{Dirac}) reduce to (we will only be considering the case $m=0$ in this paper and $a_1=1$, $a_2=1$)

\begin{equation}
\left(x+\frac{p^2 \sin(F)^2}{x}\right)F''+\frac{p^2\sin(2F)}{2x}F'^2+\left(1-\frac{p^2\sin(F)^2}{x^2}\right)F'-x\kappa_0\sin(F)-\frac{p^2\sin(2F)}{2x} \nonumber
\end{equation}
\begin{equation}
\quad+\frac{g x}{N_i^2}\left(2\cos(F)(u_1 u_2-v_1 v_2)+\sin(F)\left(-u_1^2+u_2^2+v_1^2-v_2^2\right)\right)\nonumber
\end{equation}
\begin{equation}
\quad\quad-x\;\omega^2\sin(F)\left(-\cos(F)+\frac{\sin(F)F'}{x}+\cos(F)F'^2+\sin(F)F''\right)=0,  \label{aux2}
\end{equation}
\begin{equation}
v_1^{\prime}-\frac{l v_1}{x}-g\sin(F)u_2-\left(k_1\omega+g\cos(F)\right)u_1=0,  \label{final1}
\end{equation}%
\begin{equation}
v_2^{\prime}-\frac{(l+p)v_2}{x}-g\sin(F)u_1-\left((1+k_1)\omega-g\cos(F)\right)u_2=0,  \label{final2}
\end{equation}%
\begin{equation}
u_1^{\prime}+\frac{(1+l)u_1}{x}-g\sin(F)v_2+\left(k_1\omega-g\cos(F)\right)v_1=0,  \label{final3}
\end{equation}%
\begin{equation}
u_2^{\prime}+\frac{(1+l+p)u_2}{x}-g\sin(F)v_1+\left((1+k_1)\omega+g\cos(F)\right)v_2=0.  \label{final4}
\end{equation}%
As anticipated before, our ansatz leads to a set of radial equations for the profile-functions, decoupling the angular dependence from the system. 

In the following, we choose to restrict our attention to the state characterized by $l=0$ and $p=-1$. This choice will help the comparison of the previous equations with the ones for the energy Eigenstates in section \ref{energyeigen}, in which $l=0$ and $p=-1$ correspond to the states with zero total angular momentum.  

For the Fermionic profile function it's important to check the asymptotic behavior at $x\rightarrow \infty$. Then, one can set $f=0$ and solve analytically the equations that result, giving a Fermionic behaviour of the form
\begin{equation}\label{bc1}
v_1 = u_1 \approxeq C_1 K_0\left(\sqrt{g^2-k_1^2\omega^2}x\right)
\end{equation}
\begin{equation}
v_2 = u_2 \approxeq C_2 K_0\left(\sqrt{g^2-(1+k_1)^2\omega^2}x\right),
\end{equation}

for some constant $C_i$ and where $K_0$ is a Bessel function. This means in particular that exponentially localised Fermionic states can only be found if $g^2>(1+k_1)^2\omega^2$. Note that this imposes a third and novel constraint on the possible values of $\omega$, since
\begin{equation}
    |\omega|<\big|\frac{g}{1+k_1}\big|.
\end{equation}
Similarly, the relevant boundary conditions at the origin $x\rightarrow 0$ are 
\begin{equation}\label{bc2}
v_2 = u_1 =0,\qquad  v_1' = u_2' =0.
\end{equation}Given the system (\ref{aux2})-(\ref{final4}), once chosen the integer numbers $l=0$ and $p=-1$, and the angular isospin velocity $\omega$, then there exists a value of $k_1$ for which the system is solvable within the required boundary conditions, giving smooth solutions for the profile functions. Therefore, in this sense, the resolution of the equations of motion is numerically similar to an Eigenstates problem where in our case $k_1$ plays the role of the unknown Eigenvalue.

Note that while one may think that just taking $\omega=0$ naively should bring us back to the case without isospin studied in \cite{Perapechka:2018yux}, a more subtle limit must be considered. In fact, the right procedure to reduce our system to the one of \cite{Perapechka:2018yux} is to take the limit $\omega\rightarrow 0$ together with $k_1\rightarrow \infty$, while keeping $k_1\omega = const$. Then, the energy Eigenvalue $\epsilon$ of such a state would be given by $\epsilon=-k_1\omega$. It is clear from the form of the Fermion ansatz (\ref{ansatzzf}) that in this limit the Fermion becomes an exact Eigenstate of a time-independent Hamiltonian, taking the common form $\Psi=\psi(x,\varphi)e^{-i\epsilon t}$. It was found in \cite{Perapechka:2018yux}, in the system with no isospin, that when the mass of the Fermions is zero, the Fermion equations reduce to two equations since $u_1=-v_2$ and $u_2=v_1$. However, when $\omega\neq0$ this is not the case. It is only true in the limit discussed above. The difference in shapes between the field profiles is therefore a good measure of how far one is from the limit of exact Eigenstates.\newline

We therefore wish to solve equations (\ref{aux2})-(\ref{final4}) subject to boundary conditions (\ref{bc1})-(\ref{bc2}) and the integral constraint (\ref{intcont}). These equations have to be solved numerically. For our numerical solver, we compactify the spatial domain using the change of coordinates $x=\frac{z}{1-z}$, with $z\in\left[0,1\right]$. Our solver discretises the differential equations using a second order central difference scheme. The resulting set of simultaneous equations are solved using a Newton Method. We vary the number of points on the $\left[0,1\right]$ grid we use for our discretization in accordance to the accuracy required by the scale of our solutions, we checked that all our solutions are accurate up to $\mathcal{O}\left(10^{-6}\right)$. As a further check, we always substituted our solutions back into the equations and calculated their norm, finding that the equations are satisfied to order $\mathcal{O}\left(10^{-12}\right)$. For our normalization, we always use an occupation number $N_i = 1$. The code was fully implemented in Wolfram language. \newline 

In figures \ref{fig0} and \ref{fig1} we show localised Fermionic solutions. As explained above, the solutions respect the condition  $g^2>(1+k_1)^2\omega^2$. In order to raise the back-reaction parameter $g$, we must be careful to respect the additional constraints posed on $\omega$ while making sure the numerical procedure converges on a solution. 
Figures \ref{figlargeg1} and \ref{figlargeg4} shows solutions where the backreaction parameter is increased. Moreover, in figure \ref{figlargeg1} and \ref{figlargeg4} we see solutions with nodes on the radial axis. This causes an intermediate series of zeros in the Fermion density on the radial coordinate.  \newline

For fixed values of $g$, $\kappa_0$ and $\omega$, we have not been able to find multiple branches of solutions with different $k_1$ and different number of nodes on the Fermion functions. However, we cannot exclude the existence of this possibility. In figure \ref{fig2}, we solved the system fixing the value of $k_1$ and finding $\omega$ for different $g$ at fixed $\kappa_0$. As said above and as shown in the figure, we found just one value of $\omega$ that makes the system solvable. Importantly, the point corresponding to $\omega=0$ on this plot is a point of exact coincidence between solutions to the equations of motion and Hamiltonian Eigenstates. This is obvious, since $\omega=0$ in this case coincides with the results of the no isospin study of \cite{Perapechka:2018yux}. It is a good check of our numerical procedure that the point on the $g$ axis for which $\omega=0$ is closed to the one presented in the Eigenstate analysis in figure \ref{spec} in section \ref{energyeigen}, although a different $\kappa_0$ has been used. 

\begin{figure}[h!]
	\begin{subfigure}{0.475\linewidth}
		\includegraphics[width=0.95\linewidth]{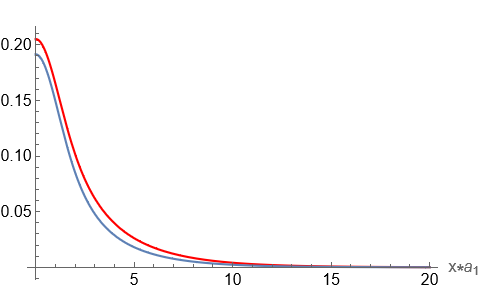}
	\caption{$v_1$ (blue) and $u_2$ (red) profile functions.}
	\end{subfigure}
	\begin{subfigure}{0.475\linewidth}
		\includegraphics[width=0.95\linewidth]{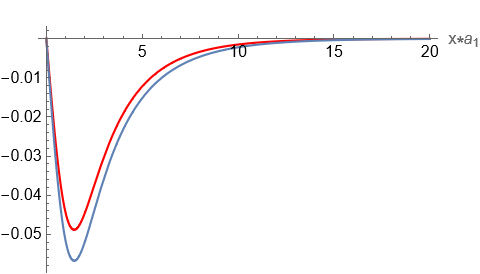}
	\caption{$v_2$ (blue) and $- u_1$ (red) profile functions.}
	\end{subfigure}
\centering
	\begin{subfigure}{0.475\linewidth}
		\includegraphics[width=0.95\linewidth]{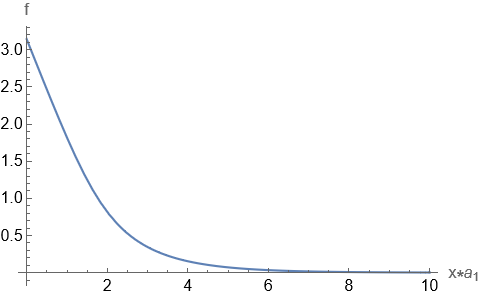}
	\caption{Baby-skyrme profile function.}
	\end{subfigure}
	\begin{subfigure}{0.475\linewidth}
		\includegraphics[width=0.95\linewidth]{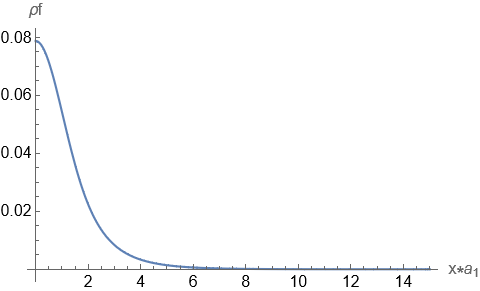}
	\caption{Fermion density}
	\end{subfigure}
	\caption{Numerical solutions for the field profiles with $m=0$.  We use $k_1=-5$, $g/a_1=-|g|/a_1=-0.4$, $\omega/a_1=-0.052$, $\kappa_0/a_1^3=0.4$, $a_1=1$, $a_2=1$, $l=0$ and $p=-1$. The plots show that $v_1 \neq u_2$ and $v_2 \neq - u_1$, indicating that we are far from the limit in which these solutions are also Eigenstates of the Hamiltonian.}
\label{fig0}
\end{figure}

\begin{figure}[h!]
	\begin{subfigure}{0.475\linewidth}
		\includegraphics[width=0.95\linewidth]{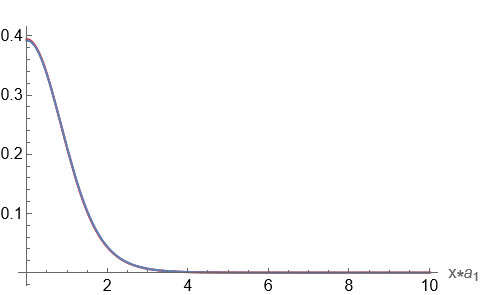}
	\caption{$v_1$ (blue) and $u_2$ (red) profile functions.}
	\end{subfigure}
	\begin{subfigure}{0.475\linewidth}
		\includegraphics[width=0.95\linewidth]{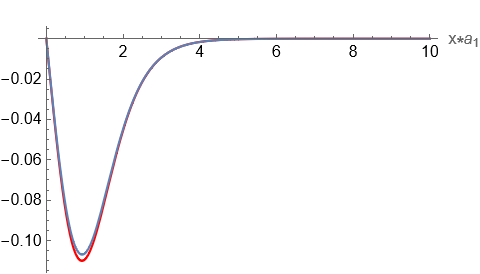}
	\caption{$v_2$ (blue) and $- u_1$ (red) profile functions.}
	\end{subfigure}
\centering
	\begin{subfigure}{0.475\linewidth}
		\includegraphics[width=0.95\linewidth]{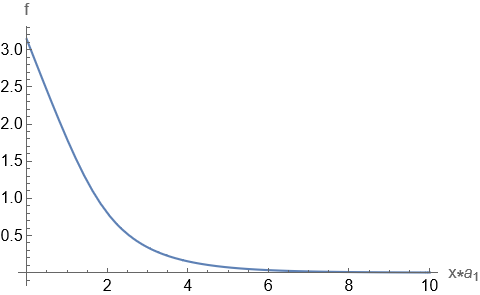}
	\caption{Baby-skyrme profile function.}
	\end{subfigure}
	\begin{subfigure}{0.475\linewidth}
		\includegraphics[width=0.95\linewidth]{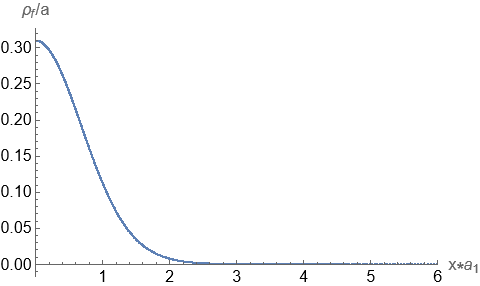}
	\caption{Fermion density}
	\end{subfigure}
	\caption{Numerical solutions for the field profiles with $m=0$.  We use $k_1=-28$, $g/a_1=-|g|/a_1=-2.0$, $\omega/a_1=-0.037$, $\kappa_0/a_1^3=0.4$, $a_1=1$, $a_2=1$, $l=0$ and $p=-1$. The plots show that $v_1 \approx u_2$ and $v_2 \approx - u_1$, indicating that we are close to the limit in which these solutions are also Eigenstates of the Hamiltonian.}
\label{fig1}
\end{figure}

\begin{figure}[h!]
	\centering
		\includegraphics[width=0.6\linewidth]{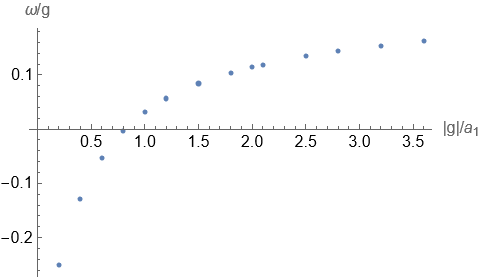}
	\caption{Values of $k_1\omega$ of solutions as a function of the baby-Skyrme coupling $g = -|g|$ with $m=0$. For these solutions $\kappa_0=0.4$, $a_1=1$, $a_2=1$ and $l=0$, $p=-1$ and $k_1=-5$.}
\label{fig2}
\end{figure}

\begin{figure}[h!]
	\begin{subfigure}{0.475\linewidth}
		\includegraphics[width=0.95\linewidth]{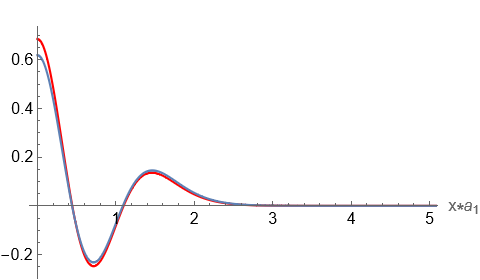}
	\caption{$v_1$ (blue) and $u_2$ (red) profile functions.}
	\end{subfigure}
	\begin{subfigure}{0.475\linewidth}
		\includegraphics[width=0.95\linewidth]{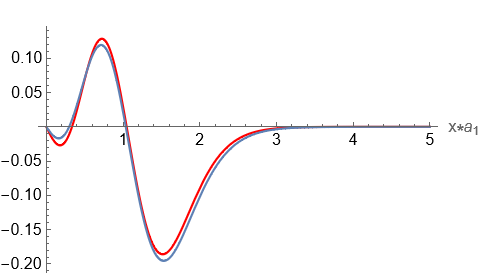}
	\caption{$v_2$ (blue) and $- u_1$ (red) profile functions.}
	\end{subfigure}
\centering
	\begin{subfigure}{0.475\linewidth}
		\includegraphics[width=0.95\linewidth]{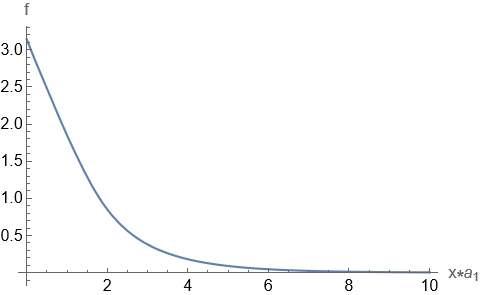}
	\caption{Baby-skyrme profile function.}
	\end{subfigure}
	\begin{subfigure}{0.475\linewidth}
		\includegraphics[width=0.95\linewidth]{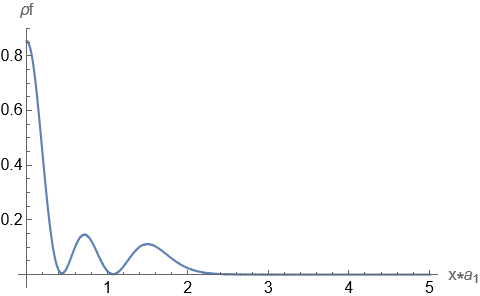}
	\caption{Fermion Density}
	\end{subfigure}

	\caption{Numerical solutions for the field profiles with $m=0$.  We use $k_1=-70$, $g/a_1=-|g|/a_1=-20.4$, $\omega/a_1=-0.284$, $\kappa_0/a_1^3=0.4$, $a_1=1$, $a_2=1$, $l=0$ and $p=-1$. The plots show that $v_1 \approx u_2$ and $v_2 \approx - u_1$, indicating that we are close to the limit in which these solutions are also Eigenstates of the Hamiltonian.}
\label{figlargeg1}
\end{figure}

\begin{figure}[h!]
	\begin{subfigure}{0.475\linewidth}
		\includegraphics[width=0.95\linewidth]{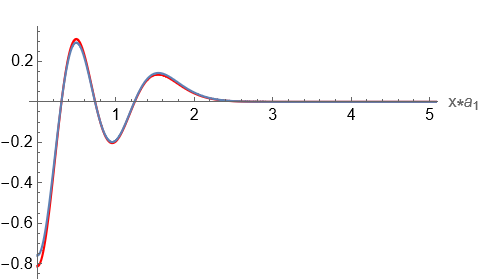}
	\caption{$v_1$ (blue) and $u_2$ (red) profile functions.}
	\end{subfigure}
	\begin{subfigure}{0.475\linewidth}
		\includegraphics[width=0.95\linewidth]{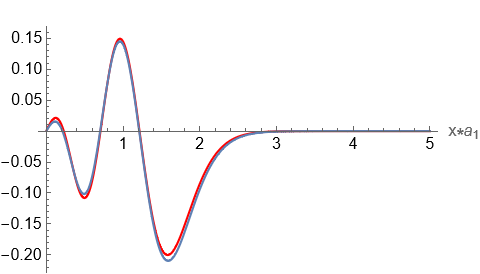}
	\caption{$v_2$ (blue) and $- u_1$ (red) profile functions.}
	\end{subfigure}
\centering
	\begin{subfigure}{0.475\linewidth}
		\includegraphics[width=0.95\linewidth]{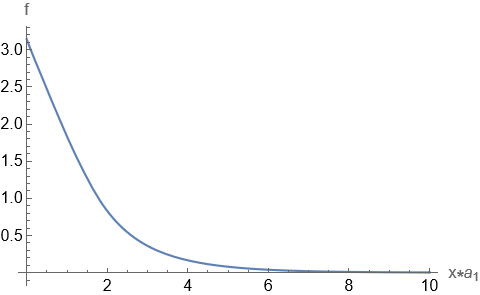}
	\caption{Baby-skyrme profile function.}
	\end{subfigure}
	\begin{subfigure}{0.475\linewidth}
		\includegraphics[width=0.95\linewidth]{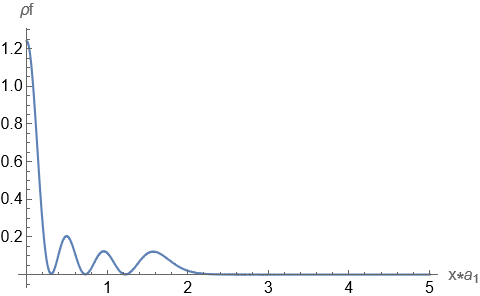}
	\caption{Fermion Density}
	\end{subfigure}

	\caption{Numerical solutions for the field profiles with $m=0$.  We use $k_1=-205$,  $g/a_1=-|g|/a_1 = -40.4$, $\omega/a_1= -0.195$, $\kappa_0/a_1^3=0.4$, $a_1=1$, $a_2=1$, $l=0$ and $p=-1$.  The plots show that $v_1 \approx u_2$ and $v_2 \approx - u_1$, indicating that we are close to the limit in which these solutions are also Eigenstates of the Hamiltonian.}
\label{figlargeg4}
\end{figure}

\subsection{Energy Eigenstates}\label{energyeigen}

In sections \ref{eoms}, we showed an example of time-evolution for a baby Skyrmion-Fermion system with non-zero isospin for the baby-Skyrmion. Using the proper ansatz (\ref{babyansatz}) and (\ref{fermionansatz1})-(\ref{ansatzzf}), we proved that a time-evolution for such a system with rotationally invariant solutions exist. Moreover, these solutions have stationary Baryonic and Fermionic density, being the profile functions $f$, $v_1$, $v_2$, $u_1$ and $u_2$ time-independent. However, as we are going to show in this section, differently from the case studied in \cite{Perapechka:2018yux} with no isospin, the stationary and rotational-invariant solutions found here are not Eigenstates of the Hamiltonian. In the case of \cite{Perapechka:2018yux}, being the Hamiltonian of the system $\mathcal{H}$ time-independent, a simple ansatz of the type $\Psi=\psi(x,\varphi)e^{-i\epsilon t}$ reduces the Fermionic equation of motion (\ref{Dirac}) to the Eigenvalue-equation $\mathcal{H}\psi=\epsilon\psi$. In that case, therefore, once the proper ansatz for $\psi$ is chosen, the Fermionic state $\Psi$ in the time-evolution can be simultaneously an energy Eigenstate and a stationary and rotational-invariant solution of the equations of motion. This procedure, however, cannot be followed in the case of isospinning baby-Skyrmion discussed in this paper. The Fermionic Hamiltonian here is time-dependent (see below) and a decomposition of the type $\Psi=\psi(x,\varphi)e^{-i\epsilon t}$ does not work to solve the equations of motion. A non-trivial time dependence in the ansatz (\ref{ansatzzf}) is indeed required to obtain a time-independent system of equations for the profile functions and thus a stationary solution. As a result, the equations (\ref{final1})-(\ref{final4}) differ from the energy-Eigenvalue equations, as we show below.

Given the Fermion Lagrangian (\ref{def}), the Fermionic Hamiltonian reads
\begin{equation}
\mathcal{H}(t)=\left[ 
\begin{array}{cccc}
-g\cos F  & -g \sin F e^{ -i\left( p\varphi
+\omega t\right)} & e^{-i\varphi}\left(\frac{x\partial _{x}-i\partial_ \varphi }{x} \right)& 0 \\ 
-g \sin F e^{ i\left( p\varphi +\omega t\right)} & g\cos F
 & 0 & e^{-i\varphi}\left(\frac{x\partial _{x}-i\partial_ \varphi }{x} \right) \\ 
 e^{i\varphi}\left(-\frac{x\partial _{x}+i\partial_ \varphi }{x} \right) & 0 & g\cos F  & g\sin F e^{ -i\left( p\varphi +\omega t\right)} \\ 
0 &  e^{i\varphi}\left(-\frac{x\partial _{x}+i\partial_ \varphi }{x} \right) & g\sin F e^{i\left(
p\varphi +\omega t\right)} & -g\cos F %
\end{array}%
\right]\,,\label{hamiltonian}
\end{equation}where the time-dependence is explicit. Since in this work we are considering a specific time-evolving system of the type (\ref{babyansatz}) and (\ref{fermionansatz1})-(\ref{ansatzzf}), the profile function $F=F(x)$ in the Hamiltonian (\ref{hamiltonian}) is the one resulting from the system (\ref{aux2})-(\ref{final4}).

Even if the Hamiltonian explicitly depends on time, at every given $t$ it is always possible to define a basis of energy Eigenstates with the corresponding Eigenvalues, i.e. 
\begin{equation}\label{eqeigen}
    \mathcal{H}(t)\Psi_n(x,\varphi,t)=\epsilon_n(t)\Psi_n(x,\varphi,t)\,.
\end{equation}Here, $\Psi_n(x,\varphi,t)$ is $n$th Eigenstate whose function is different at each time and $\epsilon_n(t)$ is the corresponding Eigenvalue that, in principle, can depend on time too. The complete set of Eigenstates for $\mathcal{H}(t)$ of equation (\ref{eqeigen}) has a precise physical meaning. Let us consider the operation of a measurement of energy for the time-evolving Fermionic state $\Psi$ resulting from eqs. (\ref{aux2})-(\ref{final4}) at a given time $t_0$. The Fermionic wave-function $\Psi$ at a given time $t_0$ can be written as a superposition of energy-Eigenstates $\Psi_n(x,\varphi,t_0)$ of the Hamiltonian $\mathcal{H}(t=t_0)$ as
\begin{equation}\label{superp}
    \ket{\Psi(t_0)}=\sum_n \ket{\Psi_n(t_0)}\bra{\Psi_n(t_0)} \ket{\Psi(t_0)}.
\end{equation}
Measuring the energy of the fermion at the time $t_0$, the wave-function $\Psi$ will collapse to a given energy-Eigenstate $\Psi_n(x,\varphi,t_0)$ with probability $|\bra{\Psi_n(t_0)}\ket{\Psi(t_0)}|^2$, giving thus a measured energy of value $\epsilon_n(t_0)$. \newline

In order to calculate the Eigenvalues of the Hamiltonian (\ref{hamiltonian}), we propose the following ansatz
\begin{equation}
\Psi_n(x,\varphi,t) =1/N_{i}\left( 
\begin{array}{c}
\tilde{v}_{1}\left( x\right) \exp \left[ i \tilde{\Omega} _1(\varphi,t) \right]
\\ 
\tilde{v}_{2}\left( x\right) \exp \left[ i\tilde{\Omega} _2(\varphi,t) \right] \\ 
\tilde{u}_{1}\left( x\right) \exp \left[ i\tilde{\Omega} _3(\varphi,t) \right] \\ 
\tilde{u}_{2}\left( x\right) \exp \left[ i\tilde{\Omega} _4(\varphi,t) \right]%
\end{array}%
\right)  \label{fermionansatzeig}
\end{equation}%
with
\begin{equation}\label{ansatzzfeig}
\begin{aligned}
&\tilde{\Omega}_1=l\; \varphi\\
& \tilde{\Omega}_2=(p+l)\;\varphi+\omega\;t\\
& \tilde{\Omega}_3=(l+1)\;\varphi\\
& \tilde{\Omega}_4=(l+p+1)\;\varphi+\omega t .
\end{aligned}
\end{equation} 
Inserting (\ref{fermionansatzeig}) and (\ref{ansatzzfeig}) into the equation (\ref{eqeigen}), we finally obtain the system:
\begin{equation}
\tilde{v}_1^{\prime}-\frac{l \tilde{v}_1}{x}-g\sin(F)\tilde{u}_2-\left(-\epsilon_n+g\cos(F)\right)\tilde{u}_1=0,  \label{final1eig}
\end{equation}%
\begin{equation}
\tilde{v}_2^{\prime}-\frac{(l+p)\tilde{v}_2}{x}-g\sin(F)\tilde{u}_1-\left(-\epsilon_n-g\cos(F)\right)\tilde{u}_2=0,  \label{final2eig}
\end{equation}%
\begin{equation}
\tilde{u}_1^{\prime}+\frac{(1+l)\tilde{u}_1}{x}-g\sin(F)\tilde{v}_2+\left(-\epsilon_n-g\cos(F)\right)\tilde{v}_1=0,  \label{final3eig}
\end{equation}%

\begin{equation}
u_2^{\prime}+\frac{(1+l+p)u_2}{x}-g\sin(F)\tilde{v}_1+\left(-\epsilon_n+g\cos(F)\right)\tilde{v}_2=0.  \label{final4eig}
\end{equation}
Note that, due to the lack of t-dependence, the Eigenvalues $\epsilon_n$ do not depend on time, i.e. $\epsilon_n(t)=\epsilon_n$.

Once the Eigenvalue equations are written, we finally compare them with the Fermionic equations (\ref{final1})-(\ref{final4}) of the time-evolution system for $\Psi$. It is now evident that, being $(1+k_1)\omega\neq k_1\omega$ for every $|\omega|>0$, there is no possibility for the stationary and rotational-invariant solution $\Psi$ to be simultaneously an Eigenstate of the energy. On the contrary, as already anticipated in the previous section, only in the limit $\omega\rightarrow 0$, $k_1\rightarrow \infty$ and $\omega k_1=const$, the time-evolving state $\Psi$ is effectively an energy-Eigenstate with Eigenvalue $\epsilon_n=-\omega k_1$. 

The ansatz (\ref{fermionansatzeig}) and (\ref{ansatzzfeig}) simplifies the form of eq. (\ref{superp}) and allows us to see explicitly the stationary fermionic solution (\ref{fermionansatz1})-(\ref{ansatzzf}) written as a superposition of energy-Eigenstates. In fact, at a given time $t_0$, the fermionic solution $\Psi$ can be written as 
\begin{equation}\label{superp2}
    \ket{\Psi(t_0)}=\sum_n \ket{\Psi_n(t_0)}\bra{\Psi_n(t_0)} \ket{\Psi(t_0)}=\sum_n e^{it_0\omega k_1}\,c_n \ket{\Psi_n(t_0)}
\end{equation}where the coefficient $c_n$ is given by
\begin{equation}
    c_n =  \int\left(\tilde{u}^*_1u_1+\tilde{u}_2^{*}u_2+\tilde{v}_1^{*}v_1+\tilde{v}_2^{*}v_2\right) d^2x,
\end{equation}
with $\tilde{u}_1$, $\tilde{u}_2$, $\tilde{v}_1$ and $\tilde{v}_2$ the (normalized) profile functions of the $n$th Eigenstates, whereas $u_1$, $u_2$, $v_1$ and $v_2$ are the (normalized) profile functions for the stationary state $\Psi$. The value $|c_n|^2$ represents the probability of finding the stationary solution $\Psi$ in the $n$th Eigenstate once the energy is measured.\\
As discussed before, in the limit $\omega\rightarrow 0$, $k_1\rightarrow \infty$ and $\omega k_1=const$, the equations for the time-evolving state $\Psi$ become equal to the Eigenstate-equations for a given $m$th energy-mode. Therefore, in this limit we reduce to $c_n=1$ for $n=m$, $c_n=0$ for $n\neq m$ and also $\Psi_n(x,\varphi,t)\rightarrow \Psi_n(x,\varphi)$ (being $\omega\rightarrow 0$). The equation (\ref{superp2}) becomes then
\begin{equation}\label{superp3}
    \ket{\Psi(t_0)}= e^{it_0\omega k_1}\, \ket{\Psi_m}\,,
\end{equation}that is the time-evolution of the $m$th energy-Eigenstate with Eigenvalue $\epsilon_m=-\omega k_1$, as said before. We emphasize again how eq. (\ref{superp3}) is exactly the time-evolving fermion ansatz proposed in \cite{Perapechka:2018yux}.

This limit will become apparent when comparing the mean energies of our solutions to the energy Eigenstates, in section 3.5 below. 

We can now explicitly solve the equation (\ref{final1eig})-(\ref{final4eig}) and thus calculate the energy Eigenstates and Eigenvalues of the Hamiltonian. Although the system (\ref{final1eig})-(\ref{final4eig}) appears functionally the same of \cite{Perapechka:2018yux}, we note that in our case the profile-function $F=F(x)$ comes from the solution of equations (\ref{aux2})-(\ref{final4}) in which, in addiction to the Fermion backreaction, also the isospin rotation plays a role in deforming the baby-Skyrmion profile. We expect, therefore, for $|\omega| >0$ to obtain a deformed spectral flow for the Eigenvalues $\epsilon_n$ compared to \cite{Perapechka:2018yux}. 

To perform an explicit calculation, it is convenient to define the total angular momentum operator \cite{Perapechka:2018yux}
\begin{equation}
    \hat{K}_3=-i\frac{\partial}{\partial \varphi }+\frac{\gamma_3}{2}+p\frac{\tau_3}{2}\,,
\end{equation}that counts together the spatial angular momentum, the spin and isospin. This operator commutes with the Hamiltonian (\ref{hamiltonian}) and thus it is possible to write a set of common Eigenstates. The ansatz (\ref{fermionansatzeig}) with (\ref{ansatzzfeig}) has exactly this property since
\begin{equation}
     \hat{K}_3\Psi_n(x,\varphi,t)=\kappa \Psi_n(x,\varphi,t)\qquad \text{with}\quad \kappa=\frac{1}{2}(1+p+2l)\,.
\end{equation}The total-momentum Eigenvalue $\kappa$ can be used to classify the energy-Eigenstates together with the index $n$ of $\epsilon_n$. Note that the operator $\hat{K}_3$ acts in the same way on the time-evolving state $\Psi$ of ansatz (\ref{fermionansatz1})-(\ref{ansatzzf}). This means that, although $\Psi$ is not an energy-Eigenstate, it still represents an Eigenstate of the total angular momentum.

To make a better comparison with \cite{Perapechka:2018yux}, we restrict our analysis about the energy-Eigenstates to the case of $\kappa=0$, and thus $l=0$ and $p=-1$. With this parameters, the four differential equations reduce to two, due to the relation $\tilde{u}_2=\tilde{v}_1$ and $\tilde{u}_1=-\tilde{v}_2$.

In figure \ref{spec}, we plot the value of $\epsilon / |g|$ as a function of $|g|$ with $g=-|g|$ for four different modes, called $ A_i $ and $ B_i $ corresponding to negative or positive energy Eigenvalues, for two values of $\omega $. Note that the mode $A_0$ for a particular choice of $g$ becomes a zero-mode, reproducing a well-known result from literature \cite{Perapechka:2018yux}.
It is clear that the presence of an isospin $\omega>0$ deforms the spectral flow, as expected. The presence (and number) of the zero modes is not affected by the iso-spin deformation, in general this should be protected by the index theorem, however the precise flow of the zero mode changes with the isospin parameter. In general, it appears switching on an isospin parameter shifts the solutions towards lower values of $g$ (the shift is small given the strict constraints on the possible values of $\omega$). In figures \ref{zoomsf}, we plot some zoomed versions of figure \ref{spec} in order to better appreciate the isospin deformation of the spectral flow.

A physical interpretation of the results of figure \ref{spec} turns out to be more subtle then one may expect. Focusing for example on the positive-energy part of the spectrum, figure \ref{zoomsf} (a), we recognize that for $g\gtrapprox 1.8$ the introduction of the isospin rotation has the effect of increasing the energy Eigenvalue, as we may expect. However, we note the opposite behaviour for low values of $g$. This result is probably due to a complicated mix of two phenomena: if from one side the isospin rotation generates an increase of the baby-Skyrmion energy and therefore a probable energy-raise of the whole system, it has also the effect of non-trivially deforming the baby-Skyrmion profile and thus the shape of the ``potential" in which the Fermion is trapped. Therefore, in order to give a complete explanation of the effect of the isospin rotation on the Fermion Eigenvalues, further specific studies are needed. We leave this for future work.

In figure \ref{eigenprofA0}, we show the profile-function for $\tilde{u}_1$ and $\tilde{u}_2$ (with $\tilde{v}_1$=$\tilde{u}_2$ and $\tilde{v}_2$=$-\tilde{u}_1$) for the the mode $A_0$ at two different values of $\omega$. Together with the profile functions, we plot also the baby-skyrmion profile and the fermionic probability density $\rho_F$ of eq. (\ref{fc11}). In this case, the profile functions of the Eigenstate $A_0$ have zero nodes and the probability density $\rho_F$ is concentrated around the origin of the plane. It is interesting to notice how the presence of the iso-spin rotation deforms the baby-skyrmion profile $F(x)$ (that we remind it is a solution of the equations of motion (\ref{aux2})-(\ref{final4})). The deformation of the profile function $F$ modifies the eigen-state equations (\ref{final1eig})-(\ref{final4eig}) and, in the end, the profile-function $\tilde{u}_1$, $\tilde{u}_2$, $\tilde{v}_1$ and $\tilde{v}_2$. In figure \ref{eigenprofA1}, we show the same panel for the mode $A_1$ at two different values of $\omega$. In this case, the fermionic profile functions have one node and again the probability density $\rho_F$ is mainly concentrated around the origin. 

\begin{figure}[h!]
	\centering
		\includegraphics[width=0.8\linewidth]{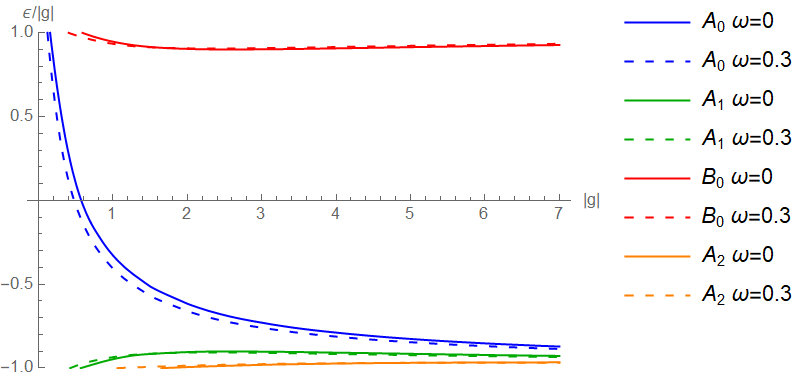}
	\caption{Plot of $\epsilon/|g|$ as a function of $|g|/a_1$ with $g=-|g|$ for the modes $A_0$, $A_1$, $A_2$ and $B_0$ for two values of isospin $\omega/a_1=0$ and $\omega/a_1=0.3$. In both cases, we used $m=0$, $\kappa_0=0.1$, $a_1=1$, $a_2=1$, $l=0$, $p=-1$.}
\label{spec}
\end{figure}

\begin{figure}[h!]
	\begin{subfigure}{0.475\linewidth}
		\includegraphics[width=0.95\linewidth]{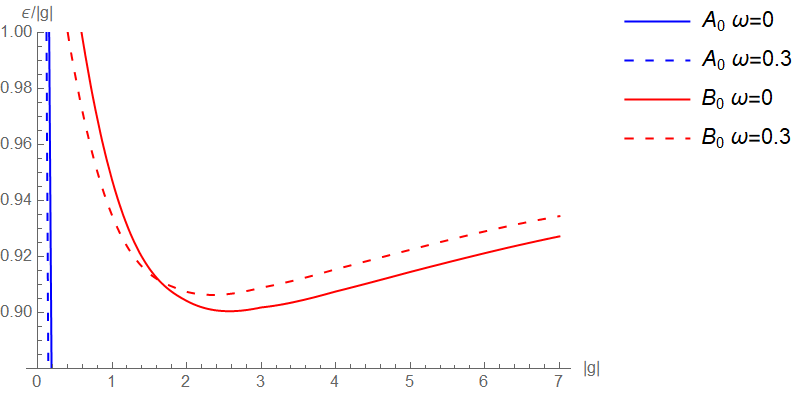}
	\caption{Zoom on the top}
	\end{subfigure}
	\begin{subfigure}{0.475\linewidth}
		\includegraphics[width=0.95\linewidth]{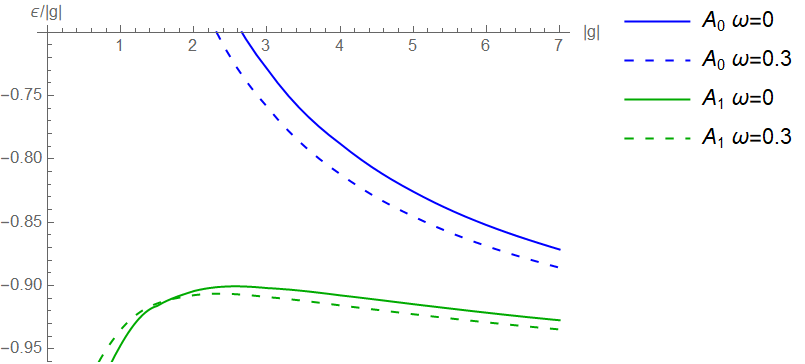}
	\caption{Zoom in the middle}
	\end{subfigure}
\centering
	\begin{subfigure}{0.475\linewidth}
		\includegraphics[width=0.95\linewidth]{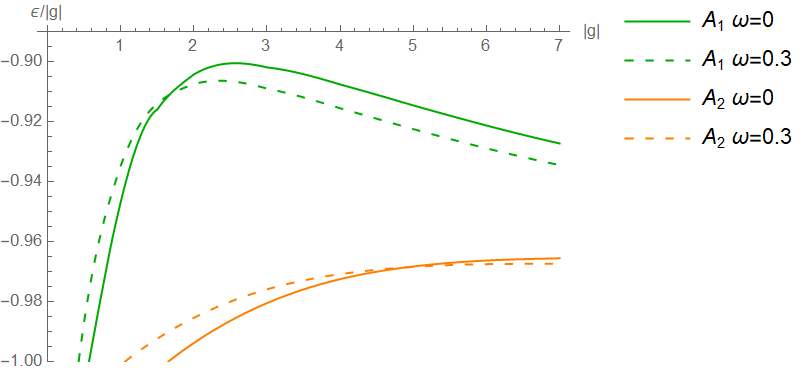}
	\caption{Zoom at the bottom}
	\end{subfigure}

	\caption{Zoomed versions of figure \ref{spec} for $\epsilon/|g|$ as a function of $|g|/a_1$ with $g=-|g|$. }
\label{zoomsf}
\end{figure}

\begin{figure}[h!]
	\begin{subfigure}{0.475\linewidth}
		\includegraphics[width=0.95\linewidth]{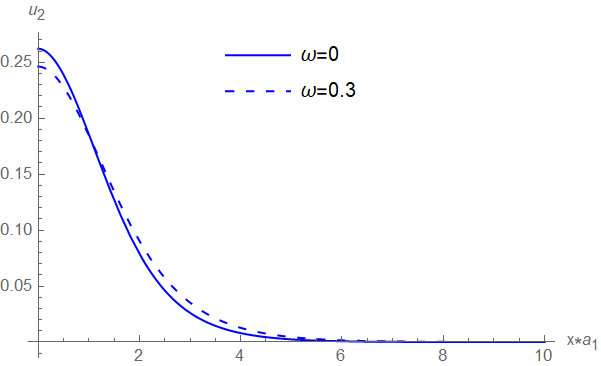}
	\caption{$u_2$ ($v_1$) profile functions.}
	\end{subfigure}
	\begin{subfigure}{0.475\linewidth}
		\includegraphics[width=0.95\linewidth]{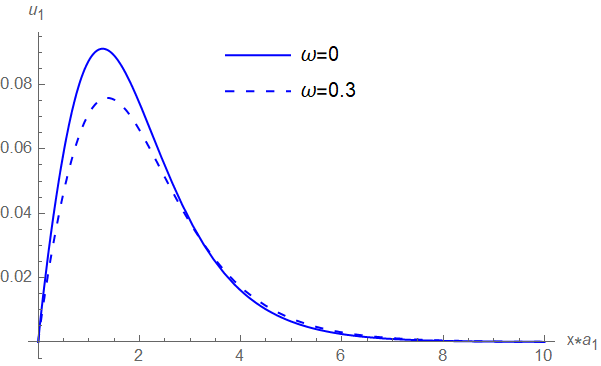}
	\caption{$ u_1$ ($-v_2$) profile functions.}
	\end{subfigure}
\centering
	\begin{subfigure}{0.475\linewidth}
		\includegraphics[width=0.95\linewidth]{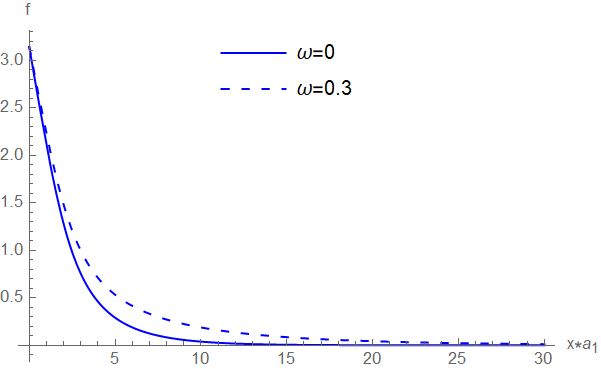}
	\caption{Baby-skyrme profile function.}
	\end{subfigure}
	\begin{subfigure}{0.475\linewidth}
		\includegraphics[width=0.95\linewidth]{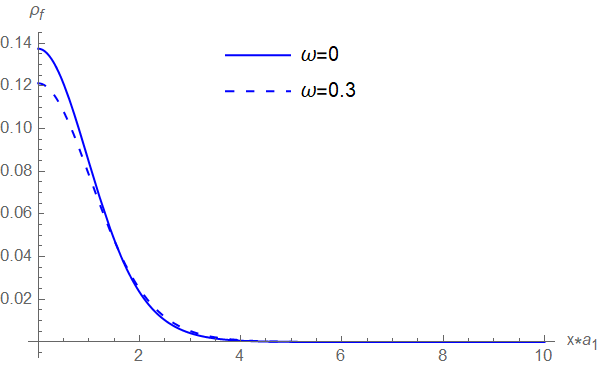}
	\caption{Fermion Density}
	\end{subfigure}

	\caption{Numerical solutions for the Eigenstate profiles of the $A_0$ mode for two different values of isospin $\omega/a_1$. We used $g/a_1=-1$, $m=0$, $\kappa_0/a_1^3=0.1$, $a_1=1$, $a_2=1$, $l=0$, $p=-1$. }
\label{eigenprofA0}
\end{figure}

\begin{figure}[h!]
	\begin{subfigure}{0.475\linewidth}
		\includegraphics[width=0.95\linewidth]{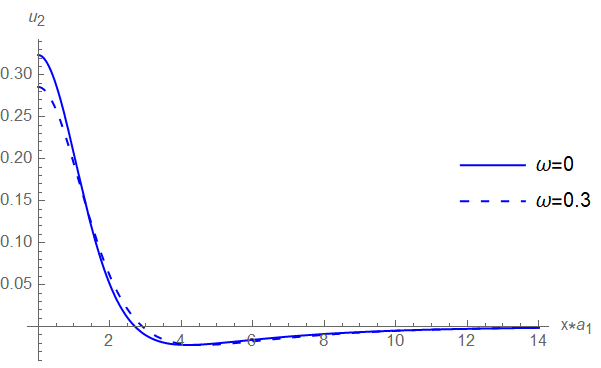}
	\caption{$u_2$ ($v_1$) profile functions.}
	\end{subfigure}
	\begin{subfigure}{0.475\linewidth}
		\includegraphics[width=0.95\linewidth]{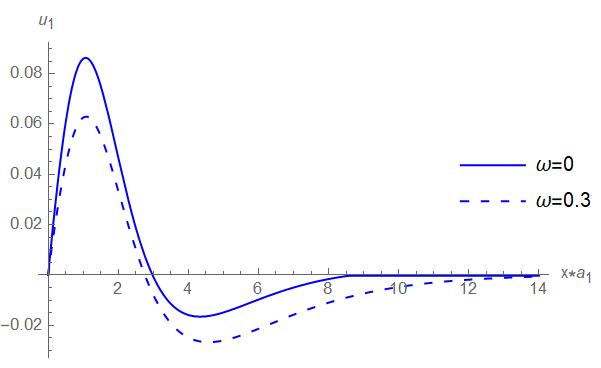}
	\caption{$ u_1$ ($-v_2$) profile functions.}
	\end{subfigure}
\centering
	\begin{subfigure}{0.475\linewidth}
		\includegraphics[width=0.95\linewidth]{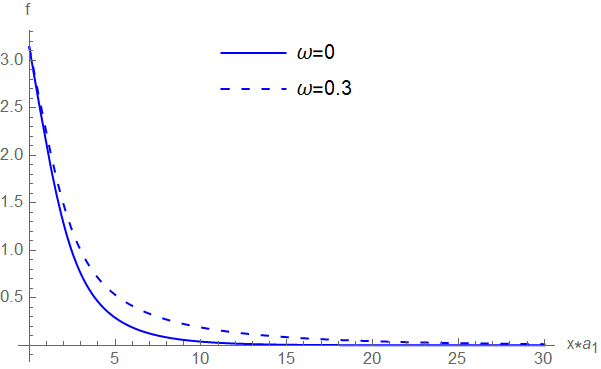}
	\caption{Baby-skyrme profile function.}
	\end{subfigure}
	\begin{subfigure}{0.475\linewidth}
		\includegraphics[width=0.95\linewidth]{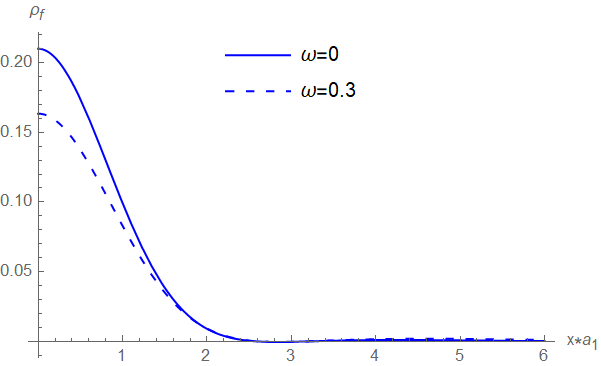}
	\caption{Fermion Density}
	\end{subfigure}

	\caption{Numerical solutions for the Eigenstate profiles of the $A_1$ mode for two different values of isospin $\omega/a_1$. We used $g/a_1=-1$, $m=0$, $\kappa_0/a_1^3=0.1$, $a_1=1$, $a_2=1$, $l=0$, $p=-1$. }
\label{eigenprofA1}
\end{figure}

\subsection{\normalsize Energy and currents}
\label{sec:energy}

{\normalsize The energy density $\rho_s$ for the baby-Skyrmion is given by the Skyrmionic component of energy tensor $T^{\mu\nu}$, this is

\begin{eqnarray}
\rho_s=T^{00}=T_{00} & =& a_{1}\left( \nabla_{0}F\nabla_{0}F+\sin^{2}{F}\left(
\nabla_{0}G\right) \left( \nabla_{0}G\right) \right)  \notag
\label{endenspre} \\
&&
-a_2\,\big\{\sin^2F[(\nabla_{0}F\nabla_{0}F)(\nabla_{\sigma}G\nabla^{%
\sigma}G)+(\nabla_{\sigma}F\nabla^{\sigma}F)(\nabla_{0}G\nabla_{0}G) 
\notag \\
&&\qquad -2(\nabla_{0}F\nabla_{0}G)(\nabla_{\sigma}F\nabla^{\sigma}G)]\big\}
\notag \\
&& -g_{00}\big\{\frac{a_1}{2}[\,\nabla_{\sigma}F\nabla^{\sigma}F+\sin^2F(%
\nabla_{\sigma}G\nabla^{\sigma}G)\,]  \notag \\
&& \qquad -\frac{a_2}{2}[\,\sin^2F((\nabla_{\sigma}F\nabla^{\sigma}F)(\nabla_{%
\sigma}G\nabla^{\sigma}G)-(\nabla_{\sigma}F\nabla^{\sigma}G)^2)\,]\\&&-\kappa _{0}\big(1-%
\vec{\Phi}\cdot \vec{n}\big)\big\}.
\end{eqnarray}

With the ansatz given by (\ref{babyansatz}), the baby skyrmion energy density reads 
\begin{equation}\label{babyen}
\rho_s = \frac{a_1\omega^2}{2}\sin^2F\big(1+\frac{a_2}{a_1}F'^2\big)+\frac{a_1}{2}F'^2+\frac{a_1p^2}{2x^2}\sin^2F\big(1+\frac{a_2}{a_1}F'^2\big)+\kappa_0(1-\cos F).
\end{equation}On the other side, the fermion energy $E_f$ can be defined only as the mean value of the Hamiltonian on the time-evolving state $\Psi$
\begin{equation}
\begin{aligned}
     E_f=\int \big[ \Psi^*\, \mathcal{H}\,\Psi\Big] \, x dx d\varphi=\frac{1}{N_i^2}\int\,\Big[&\,\left(\frac{u_1 v_1-u_2 v_2}{x}\right)+2g\sin(f)\left(u_1 u_2-v_1 v_2\right)\\
     &+g \cos(f)\left(u_1^2-u_2^2-v_1^2+v_2^2\right)\\
     &+v_1 u_1'+v_2 u_2'-u_1 v_1'-u_2 v_2'\Big]\,x dx d\varphi.
\end{aligned}
\end{equation}

In figure \ref{figEf}, we show the plots for $E_f$, the fermionic energy mean value, as a function of $\omega$ for low and large fixed $g$. We see that the solutions that we found have negative mean-energy-values, meaning they correspond (at least mainly) to superposition of modes of the negative part of the energy spectrum (these were called $A$ modes in \cite{Perapechka:2018yux} and above).

 We notice that in \ref{figEf}-right panel all the points belong to a range for which $k_1\gg 1 $. This means that, for this range of $\omega$, we are close to the limit in which these solutions are also Eigenstates of the Hamiltonian. More precisely, the different solutions of $\Psi$ consist of superpositions of energy Eigenstates well-peaked on a single mode ($A_1$, $A_2$, etc.). To emphasize this behaviour, we draw the lines corresponding to the energy-eigenvalues of the modes $A_2$ and $A_3$ (the mode $A_3$ was not considered in the spectral flow in figure \ref{spec}). As expected, in the regime $k_1\gg 1 $, the various values of $E_f$ are close to be energy-eigenvalues.  This behaviour is even more evident in the limit of $\omega$ vanishing (with $k_1$ becoming very high) that corresponds to the case analysed in \cite{Perapechka:2018yux}. This corresponds to the leftmost points in the right panel of \ref{figEf}, in which we can see that the solution mean energies are coincident with the Eigenvalues. As one increases $\omega$, one proceeds in the direction of solutions no longer being Eigenstates, therefore the solutions energies no longer coincide (this can be seen in the rightmost points of the right panel of Figure \ref{figEf}). The leftmost panel of \ref{figEf} shows instead solutions with low value of $k_1$. For all ranges of $\omega$, and for the reasons discussed above, these solutions are far from being coincident with any Eigenstate of the Hamiltonian. Having finite mean energy means that they can be decomposed in sums of energy Eigenstates, however this decomposition is not a-priori obvious and therefore one cannot assign a specific energy Eigenstate that is close to these solutions.\\
 
Another important comment concerning figure \ref{figEf}-right panel regards the jump in the value of $E_f$ at a given point of $\omega/g$. Studying the shapes of the solutions, we notice that the sudden change in the value of $E_f$ is accompanied by the addition of a node in the profile functions $u_1$, $u_2$, $v_1$ and $v_2$ of $\Psi$. This non-smooth modification with additional presence of nodes in the shape of our solutions for different values of $\omega$ and $g$ has been already shown in figures  \ref{fig0}, \ref{fig1}, \ref{figlargeg1} and \ref{figlargeg4}, where we solved the equations of motion for different parameters $\omega$ (or $k_1$) and $g$. This is expected as different energy Eigenstates of the Hamiltonian can correspond to different numbers of nodes, and thus different superpositions of Eigenstates can possess different numbers of nodes too. The fact that in figure \ref{figEf}-left panel we do not see any jump coincides with all the solutions corresponding to the different points having the same number of nodes. There remains however the question of why this sudden jump happens in the first place. It is perfectly reasonable to expect that as one adiabatically varies the value of $\omega/g$ one would stay within the space of solutions with equal numbers of nodes. The sudden jump seems therefore to be related to a global minimization of the energy, as a function of $\omega/g$. We attempted to remain within the same node sector for increasing values of $\omega/g$ by varying this vary slowly, however our numerical procedure always jumps to the new set of solutions.

\begin{figure}[h!]
	\begin{subfigure}{0.495\linewidth}
		\includegraphics[width=\linewidth]{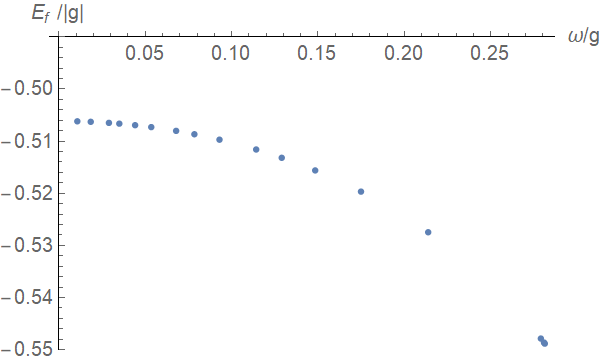}
	
	\end{subfigure}
	\begin{subfigure}{0.495\linewidth}
		\includegraphics[width=\linewidth]{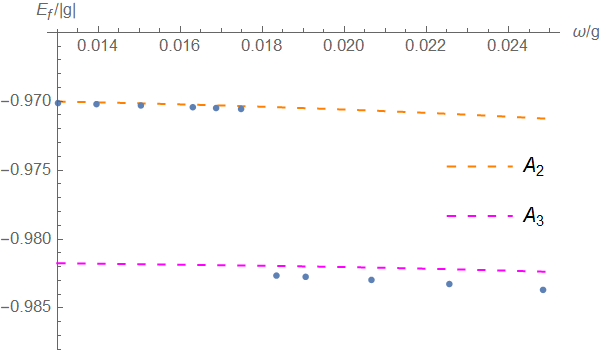}
	
	\end{subfigure}

	\caption{Numerical solutions for the integrated values of $E_f$, the mean Fermion energy, as functions of $\omega/g$ for fixed $g =- 2.0$ (left) and  $g = -20.4$ (right) with $m=0$. Moving along $\omega$ means equivalently moving along $k_1$, so that the relative interval of $k_1$ is $k_1\in[-48,-2.488]$ (left) and $k_1\in[-75,-40]$ (right). In these plots $\kappa_0/a_1^3 =0.4$, $a_1=1$, $a_2=1$, $l=0$ and $p=-1$. The dashed lines on the right figure represent the energy-eigenvalue for the modes $A_2$ and $A_3$.}
\label{figEf}
\end{figure}

\begin{figure}[h!]
	\begin{subfigure}{0.495\linewidth}
		\includegraphics[width=\linewidth]{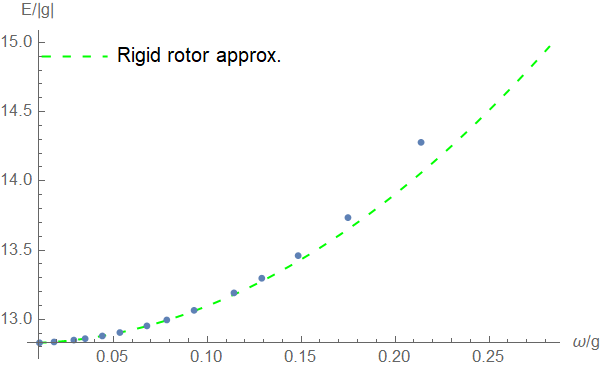}
	\caption{Total energy of solutions at fixed $g$.}
	\end{subfigure}
	\begin{subfigure}{0.495\linewidth}
		\includegraphics[width=\linewidth]{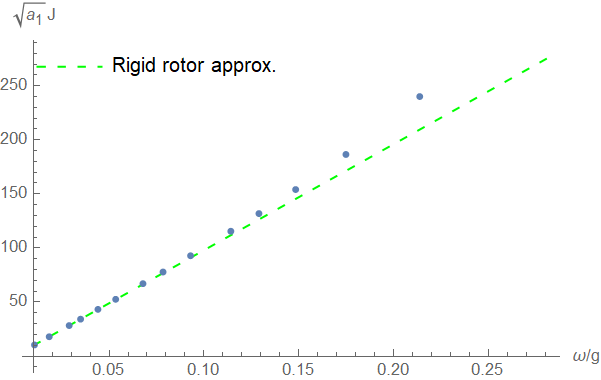}
	\caption{Total value of $J$ at fixed $g$}
	\end{subfigure}

	\caption{Numerical solutions for the integrated values of the baby Skyrmion energy $E$ and angular momentum $J$ as functions of $\omega/g$ for fixed $g/a_1=-|g|/a_1 =-2.0$ with $m=0$. In these plots $\kappa_0/a_1^3 =0.4$, $a_1=1$, $a_2=1$, $l=0$ and $p=-1$. The dashed green line represents the rigid-rotor-approximation.}
\label{fig1en}
\end{figure}

For what concerns the conserved current, the explicit form of the Skyrmion component of the global $J_{U(1)}^{\mu}$ current defined in (\ref{scc1}) that results is
\begin{eqnarray}
&&J_{s}^0= a_1 \omega\sin ^{2}F\left( 1+\frac{a_{2}}{a_{1}}F'^2\right),\\
&&J_{s}^x=0,\\
&&J_{s}^{\varphi}=\dfrac{ a_{1}p}{x^2}\sin ^{2}F\left( 1+\frac{a_{2}}{a_{1}}F'^2\right).
\end{eqnarray}
Finally, we write explicitly the form of the Fermion probability density and the probability density current defined in (\ref{fc1})
\begin{eqnarray}
&&J^{0 }_{F}=\rho_F=\frac{1}{N_i^2}\left(v_1v_1+v_2v_2+u_1u_1+u_2u_2\right),\\
&&J^x_F=0,\\
&&J^{\varphi}_F= \frac{2}{x N_i^2}\left(u_1v_1+u_2v_2\right).
\end{eqnarray}

\begin{figure}[h!]
	\begin{subfigure}{0.495\linewidth}
		\includegraphics[width=\linewidth]{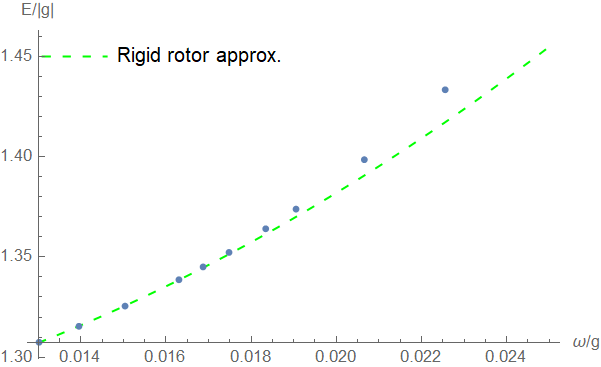}
	\caption{Total energy of solutions at fixed $g$.}
	\end{subfigure}
	\begin{subfigure}{0.495\linewidth}
		\includegraphics[width=\linewidth]{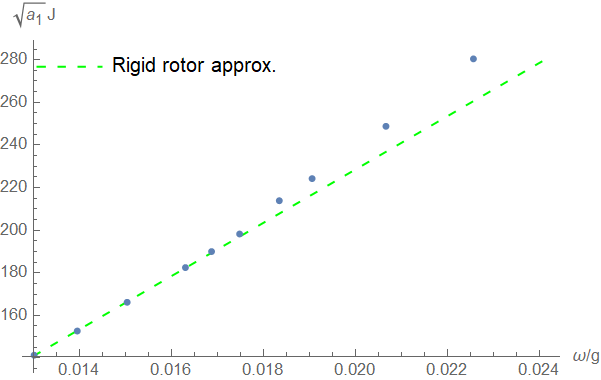}
	\caption{Total value of $J$ at fixed $g$}
	\end{subfigure}

	\caption{Numerical solutions for the integrated values of the baby Skyrmion energy $E$ and angular momentum $J$ as functions of $\omega$ for fixed $g/a_1=-|g|/a_1=-20.4$ with $m=0$. In these plots $\kappa_0/a_1^3 =0.4$, $a_1=1$, $a_2=1$, $l=0$ and $p=-1$. The dashed green line represents the rigid-rotor-approximation.}
\label{figlargeg2}
\end{figure}



\begin{figure}[h!]
	\begin{subfigure}{0.495\linewidth}
		\includegraphics[width=\linewidth]{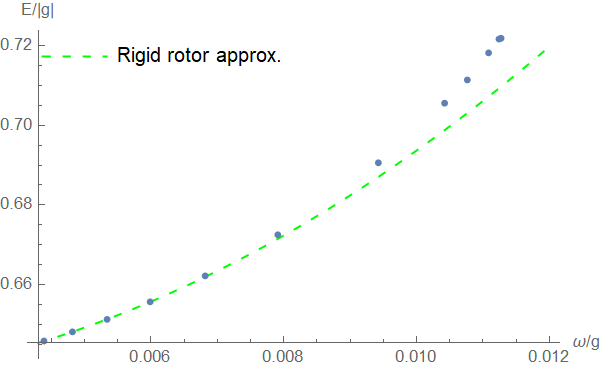}
	\caption{Total energy of solutions at fixed $g$.}
	\end{subfigure}
	\begin{subfigure}{0.495\linewidth}
		\includegraphics[width=\linewidth]{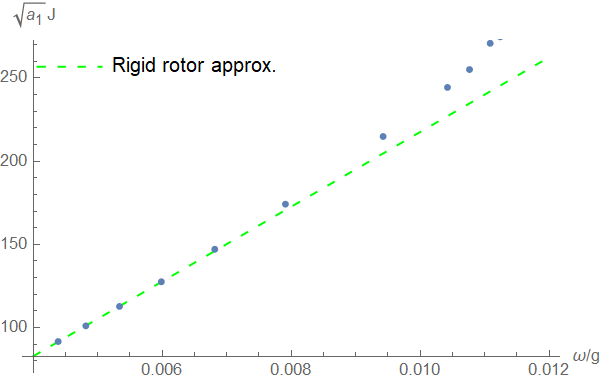}
	\caption{Total value of $J$ at fixed $g$}
	\end{subfigure}

	\caption{Numerical solutions for the integrated values of the baby Skyrmion energy $E$ and angular momentum $J$ as functions of $\omega/g$ for fixed $g/a_1=-|g|/a_1 = -40.4$ with $m=0$. In these plots $\kappa_0/a_1^3 =0.4$, $a_1=1$, $a_2=1$, $l=0$ and $p=-1$. The dashed green line represents the rigid-rotor-approximation.}
\label{figlargeg5}
\end{figure}

In figures \ref{fig1en}, \ref{figlargeg2} and \ref{figlargeg5}, we show the plots of the baby-skyrmion energy (\ref{babyen}) and the effective $J$ defined in equation (\ref{effJ}) for different values of the coupling $g$. In all these plots, we compare our results with the so-called rigid rotor approximation, i.e. the approximation in which the energy (and angular momentum) of the isospinning baby skyrmion is calculated using the soliton-profile undeformed by the iso-rotation. In that approximation, the energy and the angular momentum would increase quadratically and linearly in $\omega$, as shown in eqs. (\ref{babyen}) and (\ref{effJ}). As shown in figures \ref{fig1en}, \ref{figlargeg2} and \ref{figlargeg5}, the approximation works only for small values of $\omega$, as expected.

\section{ Conclusions}\label{sectconc}

In this paper we investigated the coupling between baby-Skyrmions and Fermions in the presence of internal isospin. We restricted our analysis to the case of rotationally symmetric $B=1$ solutions and included the Fermion backreaction on the baby-Skyrme solution. We found that rotational symmetry is consistent with a Fermionic ansatz in which Fermionic solutions to the Fermion-baby-Skyrmion equations of motion are not generically also Eigenstates of the Fermion Hamiltonian. These solutions do become Eigenstates in a particular limit in which $\omega\rightarrow 0$ but $-k_1\omega = \epsilon$ is held fixed. In general, however, we also solved for the Eigenstates of the time-dependent Hamiltonian. Solutions of the equations of motion can be written as superposition of the energy-Eigenstates and thus, once one measures the energy of the system, the fermionic wave-function will collapse to one of these Eigenstates with a given probability.
\newline
As discussed in this paper, rotationally invariant isospinning baby-Skyrmions are known to suffer from strict stability constraints on the possible values of the isospin parameter $\omega$. We found that in order to localise Fermionic solutions on them, a further constraint on $\omega$ must be imposed, coming from the behaviour of the Fermion functions at large radial values. Within these constraints, we found localised solutions for the Fermion coupled to the baby-Skyrmion, for small and large values of the coupling parameter $g$. As expected from previous studies \cite{Perapechka:2018yux}, we find a tower of energy states (both positive and negative) and a single zero mode whose precise location on the $g$ line depends on the isospin parameter $\omega$. We found numerically that increasing $\omega$ has the effect of bringing the zero mode towards lower values of $g$. Furthermore, plots of the mean energy of our Fermionic solutions (see figure \ref{figEf}) show that, in the case of high $g$, adiabatic variations of the isospin parameter lead to jumps in the energy mean value corresponding to rapid changes in the types of solutions found, characterised by different number of nodes they have in the radial direction. \newline
It is well-known that higher winding number rotationally invariant isospinning baby-Skyrmions are in general unstable towards decay into lower winding number non-rotationally invariant configurations \cite{Halavanau:2013vsa}. It is an interesting avenue of research to relax the rotationally invariant constraint on the system imposed in this paper and solve the full 2D PDE system of equations to investigate if and how the isospinning localised Fermionic solutions stabilise the Skyrmion solution.

\section*{\protect\normalsize Acknowledgments}

{\normalsize  G.T is funded by a Fondecyt grant number 1200025.}

{\normalsize \bigskip }

\end{document}